# Axial Resolution Post-processing Engineering in Fresnel Incoherent Correlation Holography

Shivasubramanian Gopinath,[1,*] Joseph Rosen,[2] and Vijayakumar Anand[1,3]

[1]*Institute of Physics, University of Tartu, W. Ostwaldi 1, 50411 Tartu, Estonia*
[2]*School of Electrical and Computer Engineering, Ben Gurion University of the Negev, P.O. Box 653, Beer-Sheva 8410501, Israel*
[3]*Optical Sciences Center, Swinburne University of Technology, Hawthorn, Melbourne, VIC 3122, Australia*
*\*shivasubramanian.gopinath@ut.ee*

**Abstract:** Fresnel incoherent correlation holography (FINCH) is a self-interference-based incoherent digital holography method. In FINCH, light from an object point is split into two beams, modulated differently using two lenses with different focal distances, and creates a self-interference hologram. At least three phase-shifted holograms are recorded and synthesized into a complex hologram, which reconstructs the object image without twin image and bias noises. Compared with conventional imaging, FINCH exhibits a longer depth of focus (DOF) and higher lateral resolution. In this study, we propose and demonstrate a new method termed post-engineering of axial resolution in FINCH (PEAR-FINCH), which enables post-recording DOF engineering for the first time. In PEAR-FINCH, a library of FINCH holograms catalogued with unique axial characteristics, DOF, and focus location is recorded by changing the focal distance of one of the diffractive lenses. Selected holograms from this library are combined to engineer new axial characteristics not achievable in FINCH. A two-step reconstruction, involving numerical back-propagation and deconvolution with a point spread hologram, is implemented. Experiments with multiplane objects having large axial separations confirm that PEAR-FINCH achieves a substantially extended DOF compared with direct imaging and FINCH. PEAR-FINCH will be promising for applications in biomedical imaging, holography, and fluorescence microscopy.

## 1. Introduction

Incoherent digital holography (IDH) has emerged as a powerful 3D imaging technique for the reconstruction of incoherently illuminated scenes or self-luminous objects [1-4]. Unlike coherent holography, some IDHs use the self-interference principle: light derived from an object point is split into two waves, each is differently modulated, and both are interfered with, as opposed to interference between the object wave and reference wave in coherent holography, to record and reconstruct the 3D image of an object. One of the most prominent IDH techniques is Fresnel incoherent correlation holography (FINCH), which was introduced in 2007 [5-7]. FINCH uses an active optical device, typically a spatial light modulator (SLM), to modulate the light from an object in two different on-axis channels. At the SLM, two object waves are generated for every object point that interfere with each other to generate a self-interference object hologram (OH). At least three-phase-shifted OHs are recorded and then digitally superposed to obtain a complex hologram. This complex hologram can reconstruct the object by Fresnel backpropagation without any twin image or bias noise. Compared with direct incoherent imaging systems with the same numerical aperture (NA), FINCH has superior lateral resolution (LR) and a greater depth of focus (DOF) [8]. The above characteristics, in addition to scanning-free implementation, make FINCH desirable for various applications compared with both direct imaging (DI) systems and scanning holography methods [9-12].

The original work on FINCH faced multiple challenges: a low signal-to-noise ratio (SNR), low temporal resolution, and low axial resolution (AR) [5]. To overcome the above limitations, different techniques have been developed. To improve the SNR of FINCH, a different approach for generating two object waves using polarization multiplexing was developed [13]. In this approach, a polarizer oriented at 45° with respect to the active axis of the SLM is introduced between the object and the SLM. The above configuration allows only approximately half of the object wave's intensity to be modulated by the mask displayed on the SLM, whereas the remaining part remains unmodulated. A second polarizer placed after the SLM is also aligned at 45° with respect to the active axis of the SLM, enabling interference between the modulated and unmodulated object waves at the image sensor. This configuration avoids the reconstruction noise generated due to the scattering of light by the randomly multiplexed diffractive lenses in the original FINCH configuration. However, the improvement in the reconstruction comes at the cost of reduced light throughput owing to the implementation of two polarizers instead of one. Alternative approaches have been developed to improve the light throughput in FINCH using a geometric phase lens [14] and scattering-suppressed multiplexed diffractive lenses designed using advanced phase-retrieval algorithms [15]. The temporal resolution of FINCH was improved by spatial and polarization multiplexing approaches [16-19]. In another approach, FINCH was implemented in the framework of coded aperture imaging (CAI), where a point spread hologram (PSH) library was recorded for different distances once as a calibration step and stored on a computer. For 3D imaging of objects, only one FINCH hologram of the object was recorded and processed with the PSH library using one of the deconvolution algorithms [20-29]. The axial resolution in FINCH was improved by FINCH implemented as CAI [28, 29], and sectioning capability was introduced by a dual SLM configuration of FINCH [30, 31]. With all the above developments in FINCH, it is possible to achieve 3D imaging with a super LR, a single camera shot, and a high SNR on par with DI methods.

In all imaging systems, including FINCH, imaging characteristics are dependent upon the physical parameters of the imaging systems. In recent years, interest in achieving new limits and ranges of imaging characteristics beyond the limits set by the physical parameters has increased. Numerous superresolution techniques, such as structured illumination [32], STED [33], and STORM [34], have been developed to overcome the LR limits set by the NA. In FINCH, the LR was improved beyond the limits defined by the NA using structured illumination [35] and scattering [36]. In another study, the field of view was extended beyond the limit set by the area of the image sensor [37]. Sensing color with a monochrome camera became possible using a point spread function library recorded for different wavelengths [38, 39]. In all imaging systems, including FINCH, different imaging characteristics are intertwined, so changing one characteristic inevitably changes the others. AR and LR are considered the most important imaging characteristics that determine the spatial discriminative capability along the longitudinal and transverse directions, respectively. Both AR and LR are fundamentally linked to the system's NA. As a result, modifying the AR typically induces a corresponding change in the LR and vice versa [40]. This connection of AR and LR with NA makes it very challenging and almost impossible to selectively modify AR without affecting LR. Several hybrid imaging systems have been developed by combining the architectures of FINCH [5] and coded aperture correlation holography (COACH) [41]. The first hybridization approach enabled the tunability of AR and LR within the corresponding resolution boundaries defined for FINCH and COACH [42]. The second hybridization approach, called coded aperture with FINCH intensity responses, was introduced to achieve the AR of COACH and the LR of FINCH [43]. Importantly, both hybridization approaches lack the ability to independently tune the AR, as any modification in the AR inevitably affects the LR. To address the interdependency problem between AR and LR, various CAI methods have been developed in recent years. One such method introduces a coded phase mask design that enables the modulation of the AR without affecting the LR [44]. Some approaches use two properties, namely, the long DOF of quasi-nondiffracting beams, such as Bessel, Airy, and self-rotating beams, and randomness to control the AR independent of the LR [45-47]. When a single-beam quasi-nondiffracting beam is generated, the axial correlation length is quite long, resulting in a long DOF. However, when the uncertainty is increased by simultaneously generating multiple quasi–nondiffracting beams propagating along different directions, the axial correlation length decreases, resulting in a lower DOF. Nevertheless, the lateral correlation length is given by the NA. Therefore, it is possible to decouple AR and LR and tune AR independent of LR. Although these methods successfully decouple AR from LR, they have certain limitations, such as reliance on advanced coded phase masks [44] and the presence of speckle noise in the reconstructed images [45-47]. Most notably, these techniques support AR tuning only at the time of recording and are not capable of tuning post-recording, thereby limiting their flexibility in imaging applications. The CAI method discussed in [44] has the capability of post-recording AR tuning by remodeling its optical configuration, but it has not been demonstrated. For the first time, a hybrid imaging method called the incoherent hybrid imaging system (INCHIS) was recently introduced to enable post-recording modulation of the AR without affecting the LR [48]. INCHIS requires only two camera shots of the scene using a refractive lens and an axicon. These two intensity recordings are subsequently combined with different weights and processed using the combined point spread functions library at different depths of the lens and axicon. By controlling the weights of the intensity recordings, the AR is modulated without affecting the LR between the AR limits of the lens and the axicon post-recording. Another CAI method called post-ensemble generation with Airy beams for spatial and spectral switching (PEGASASS) was developed to control AR and spectral resolution (SR) simultaneously without affecting LR post-recording using a monochrome camera [49]. In PEGASASS, multiple camera shots of an object scene are recorded using unique cubic phase masks, and the resulting object intensity distributions are combined with different strengths and processed with the point spread functions library recorded at different depths and wavelengths. By controlling the number and strength of the distributions, the AR and SR can be controlled simultaneously without affecting the LR. A CAI-based technique for 3D imaging that enables the reconstruction of only a specific transverse plane or multiple planes at a time from the same single-shot recorded pattern, according to the user's wishes, and by postprocessing this pattern, was demonstrated in [50].

Over the years, numerous imaging methods have been developed within the CAI framework to engineer AR, both in real-time and post-recording. However, FINCH stands apart as a robust and well-established IDH method that inherently provides high DOF, making it highly suitable for a wide range of applications, including biomedical and microscopic imaging. Despite its wide adoption and demonstrated advantages, very few studies to date have addressed the engineering of the DOF for FINCH systems. Recent efforts have sought to extend the DOF of IDH systems, and they can also be applied to FINCH. Nobukawa et al. [51] proposed a bimodal IDH method capable of switching between 3D imaging and quasi-infinite depth-of-field (QIDOF) imaging by varying the phase mask displayed on the SLM. Although their approach was not implemented in a FINCH configuration, their method can be applied to FINCH-like systems to extend the DOF. While this approach demonstrates extended DOF imaging, it faces several key limitations: (i) the DOF is fixed during the time of recording and cannot be engineered post-recording, and (ii) it allows either 3D or QIDOF imaging mode in a single acquisition, but not both, simultaneously.

In this work, we propose and demonstrate for the first time a novel technique called post-engineering of axial resolution in Fresnel incoherent correlation holography (PEAR-FINCH), which enables extending the DOF in FINCH post-recording, overcoming all the above challenges. In PEAR-FINCH, a library of OHs is recorded with different focusing relationships between the two interfering object waves, each with a unique reconstruction distance and DOF. A synthetic OH (SOH) is generated by combining the recorded OHs from the library and is reconstructed twice – once using the regular numerical back propagation and another using a deconvolution method with synthetic PSH (SPSH), which is obtained by combining the recorded PSHs from the library. It is noted that this is different from the previous studies on engineering axial resolution post-recording, as in these cases, the point spread function library is recorded corresponding to all planes [48,49]. In PEAR-FINCH, the PSH library is recorded at only one of the planes, but with different self-interference conditions obtained by changing the focal length of the phase mask displayed on the SLM. The above process reconstructs the object information over a longer DOF than that of FINCH. This study demonstrates the applicability of PEAR-FINCH under diffusive illumination conditions and might be suitable for real-world environments. The manuscript consists of four sections. The methodology of PEAR-FINCH is presented in the following section. The experimental analysis and results of PEAR-FINCH in comparison with those of FINCH and DI are presented in the third section. The summary and conclusion are presented in the final section. The simulation studies and the multiplane regular imaging results are presented in the supplementary document sections 1 and 2.

## 2. Methodology

The optical configurations of FINCH and PEAR-FINCH are shown in Figures 1(a) and 1(b), respectively. The axial characteristics of FINCH, PEAR-FINCH, and DI are compared in Figure 1(c). Light from an object point with an amplitude of $\sqrt{I_o}$ is incident on a bifocal lens located at a distance of $z_s$. The bifocal lens consists of two lenses with focal distances $f_1$ and $f_2$, resulting in two images. The complex transparency of the bifocal lens is given as $\exp(i\Phi_{BFL}) = \exp\left(-i\frac{\pi r^2}{\lambda f_1}\right) + \exp\left(-i\frac{\pi r^2}{\lambda f_2}\right)$, where $r = \sqrt{x^2 + y^2}$. The complex amplitude incident on the bifocal lens is given as $C_1\sqrt{I_o}Q(1/z_s)$, where $Q(b) = \exp(i\pi b r^2/\lambda)$, and $C_1$ is a complex constant.

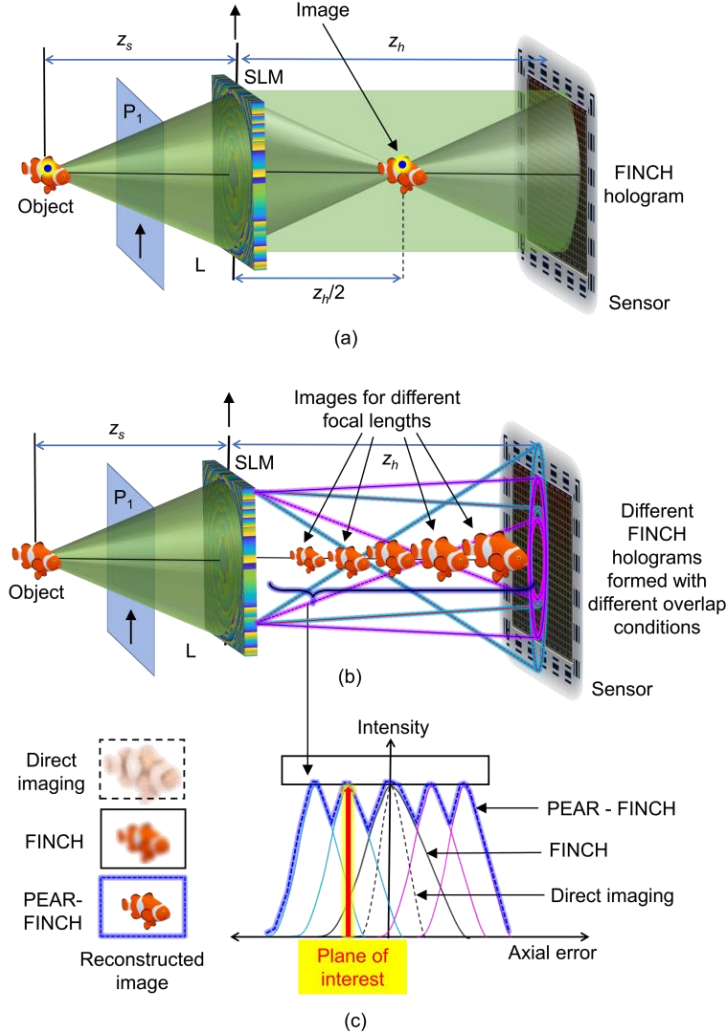

**Figure 1.** Conceptual illustration of PEAR-FINCH. (a) Optical configuration of FINCH. (b) Optical configuration of PEAR-FINCH. (c) Comparison of the axial characteristics of FINCH, PEAR-FINCH, and DI. The axial imaging responses for DI (dashed black line), FINCH (solid black line), and PEAR-FINCH (dashed black line with blue glow) are shown. At the plane of interest, DI yields the weakest and most blurred result, as it lies outside its axial response curve. FINCH provides a relatively stronger and more focused image than DI does, since the plane of interest falls within its axial curve, although not at the peak response. PEAR-FINCH enhances imaging by stitching together axial distributions from multiple FINCH focusing conditions, extending the effective depth of focus indicated by the dotted blue-glow line. This results in a highly focused image at the plane of interest, which now aligns with a region of maximal response. The cyan and pink curves represent less optimal focusing regions of FINCH on either side of the optimal condition.

The complex amplitude after the bifocal lens is given as $C_2\sqrt{I_o}Q(1/z_s)\exp(i\Phi_{BFL})$, where $C_2$ is a complex constant. The intensity $I_{PSH}$ obtained at the sensor plane located at a distance of $z_h$ is the PSH, which is expressed as

$$I_{PSH}(\bar{r}_0; z_s) = \left| C_2\sqrt{I_o}Q\left(\frac{1}{z_1}\right)\exp(i\Phi_{BFL}) \otimes Q\left(\frac{1}{z_h}\right) \right|^2$$

$$=C+Q\left(\frac{1}{z_r}\right)+Q\left(\frac{-1}{z_r}\right), \tag{1}$$

where $C$ is a real positive constant and '$\otimes$' is a 2D convolutional operator. The 3D location information of a single point is present in the distribution of $I_{PSF}$. A multipoint object consisting of $N$ points can be expressed as

$$O(\bar{r}) = \sum_{j=1}^{N} a_j \delta(\bar{r}-\bar{r}_j), \tag{2}$$

where $a_j$ is the intensity of the $j$-th point of the object $O$. FINCH is a linear, shift-invariant system, and therefore, OH is given as

$$I_{OH} = I_{PSH} \otimes O. \tag{3}$$

Equation (3) can be modified based on Eqs. (1)–(3) as

$$I_{OH}(\bar{r}_0;z_s) = \sum_{j=1}^{N} a_j I_{PSH}\left(\bar{r}_0 - \frac{z_h}{z_s}\bar{r}_j;z_s\right). \tag{4}$$

Since FINCH in Figure 1 is an in-line holography configuration, to reconstruct the object's image, it is necessary to remove the twin image and bias terms present in the hologram. This is achieved by a phase-shifting process: three holograms with different phase shifts between the interfering beams $\theta = 0, \frac{2\pi}{3}, \frac{4\pi}{3}$ are recorded and combined to remove the twin image and bias terms from the holograms. This is achieved by real-time changing of the phase of the bifocal lens $\exp(i\Phi_{BFL}) = \exp\left(-i\frac{\pi r^2}{\lambda f_1}\right) + \exp\left(-i\frac{\pi r^2}{\lambda f_2}+\theta\right)$. The complex OH can be written as $H_{OH} = I_{OH}(\theta=0)(\exp[-i4\pi/3]-\exp[-i2\pi/3]) + I_{OH}\left(\theta=\frac{2\pi}{3}\right)(1-\exp[-i4\pi/3]) + I_{OH}\left(\theta=\frac{4\pi}{3}\right)(\exp[-i2\pi/3]-1)$. This complex hologram $H_{OH}$ does not contain the twin image and bias noises when numerically backpropagated inside the computer to reconstruct the object. Moreover, for a 2D object located at a distance $z_s$ from the system, $H_{OH}$ is a series of shifted quadratic functions of the form $\sum_j a_j \delta(\bar{r} - z_h\bar{r}_j/z_s)Q(-1/z_r)$. According to the optical configuration of Figure 1, $f_1 = z_s$ and $\frac{1}{f_2} = \frac{1}{z_s} + \frac{2}{z_h}$, so the reconstruction distance is $z_r = z_h/2$. The reconstructed image is therefore given as

$$I_R = \left|H_{OH} \otimes Q\left(\frac{1}{z_r}\right)\right| = \left|\sum_j a_j \delta\left(\bar{r} - \frac{z_h}{z_s}\bar{r}_j\right) Q\left(-\frac{1}{z_r}\right) \otimes Q\left(\frac{1}{z_r}\right)\right| = O\left(\frac{z_h}{z_s}\bar{r}\right). \tag{5}$$

The DOF of FINCH can be described as the allowed tolerance around the calculated value of $z_r$, where the image is reconstructed with a relatively high lateral resolution. FINCH, unlike DI, has a longer DOF and a higher LR at the optimal configuration of perfect overlap between the two interfering beams on the sensor plane [52]. When the optical configuration is modified, i.e., when $f_1 \neq z_s$ and $\frac{1}{f_2} \neq \frac{1}{z_s} + \frac{2}{z_h}$, the lateral resolution and DOF decrease. In the above cases, the magnification of the reconstructed image is also different.

In this study, we propose a computational stitching approach, PEAR-FINCH, involving the summing of the FINCH holograms recorded under different conditions given by the different values of $f_2$ to extend the DOF of FINCH postrecording on demand. In PEAR-FINCH, the SOH generated by summing the complex holograms is given as

$$H_{SOH} = \sum_m H_{OH_m}, \tag{6}$$

where $m$ is the index for the hologram number for a particular optical configuration. The $H_{SOH}$ has a long DOF, as indicated in Figure 1, depending upon the number of summed holograms. The above $H_{SOH}$ will generate a reconstruction given as,

$$I_{R1} = |H_{SOH} \otimes Q(1/z_r)| = \left|\sum_m H_{OH_m} \otimes Q(1/z_r)\right| = \left|\sum_m \sum_j a_{m,j} H_{PSH_m} \otimes Q(1/z_r)\right|$$
$$\cong \left|\sum_j a_j \delta\left(\bar{r} - \frac{z_h}{z_s}\bar{r}_j\right) Q(-1/z_r) \otimes Q(1/z_r)\right| + NS(\bar{r}) = \sum_j a_j \delta\left(\bar{r} - \frac{z_h}{z_s}\bar{r}_j\right) + NS(\bar{r})$$
$$= O\left(\frac{z_h}{z_s}\bar{r}\right) + NS(\bar{r}), \qquad (7)$$

where $NS(\bar{r})$ is a positive noise distribution created from convolutions between $Q(1/z_r)$ and laterally shifted quadratic phase functions $Q(1/z_{\tilde{r}})$s in which $z_{\tilde{r}} \neq z_r$. Consequently, there are sharp and blurred reconstructions, which are combined, resulting in a lower SNR. This result can be improved by a second step of processing involving the SPSH, which is expressed as

$$I_{R2} = |(H_{SOH} \otimes Q(1/z_r)) \circledast (H_{SPSH} \otimes Q(1/z_r))|, \qquad (8)$$

where $H_{SPSH} = \sum_m H_{PSH_m}$, and $\circledast$ is a reconstruction operator of one of the reconstruction methods, such as nonlinear reconstruction (NLR), the Lucy–Richardson–Rosen algorithm (LRRA), and incoherent nonlinear deconvolution with an iterative algorithm (INDIA) [22, 24, 25]. Under the assumption that a nonlinear correlation can be approximated by a linear correlation, Eq. (8) can be written as

$$I_{R2} = \left|\left(\sum_m H_{OH_m} \otimes Q(1/z_r)\right) \circledast \left(\sum_m H_{PSH_m} \otimes Q(1/z_r)\right)\right|$$
$$\cong \left|\left(\sum_m H_{OH_m} \circledast \sum_m H_{PSH_m}\right) \otimes [Q(1/z_r) \circledast Q(1/z_r)]\right|$$
$$= \left|\left(\sum_m H_{OH_m} \circledast H_{PSH_m} + \sum_m \sum_{n \neq m} H_{OH_m} \circledast H_{PSH_n}\right) \otimes \delta(\bar{r})\right|$$
$$= \left|\sum_m \sum_j a_{m,j} H_{PSH_m} \circledast H_{PSH_m}\right| + NS'(\bar{r})$$
$$\cong \sum_m \sum_j a_{m,j} \delta\left(\bar{r} - \frac{z_h}{z_s}\bar{r}_j\right) + NS'(\bar{r})$$
$$\cong M \cdot O\left(\frac{z_h}{z_s}\bar{r}\right) + NS'(\bar{r}), \qquad (9)$$

where this time, $\boldsymbol{NS'(\bar{r})}$ is a positive noise distribution created from cross-correlations between object holograms and non-corresponding PSHs. Ignoring the scale differences between the various reconstructed images, Eq. (9) indicates that the reconstructed image is magnified by a factor of $M$, whereas the noise term $\boldsymbol{NS'(\bar{r})}$ remains approximately the same as in Eq. (7), because the sums in Eqs. (7) and (9) are made over random complex functions distributed uniformly over $2\pi$ phase.

## 3. Experimental analysis

The schematic and a snapshot of the optical experimental configuration are shown in Figures 2(a) and 2(b). The configuration includes the following optical components from Thorlabs: (1) M660L4 light-emitting diode (LED) with 940 mW power, operating at a wavelength of $\lambda = 660$ nm, and a bandwidth of $\Delta\lambda = 20$ nm; (2) ID50/M standard iris, Ø50.0 mm max aperture; (3) DG05-220 grit - Ø1/2" N-BK7 ground glass diffuser; (4) N-BK7 Ø1" biconvex lens refractive lens with a focal length ($f = 7.5$ cm); (5) ID50/M standard iris, Ø50.0 mm max aperture; (6) Ø1" linear polarizer with N-BK7 windows; (7) ID20/M standard iris, Ø20.0 mm max aperture; (8) N-BK7 Ø1" biconvex lens refractive lens with a focal length ($f = 5$ cm); (9) Ø1" pinhole of size (50 μm) or test object (digit '2' with a line width of 55.68 μm or digit '4' with a line width of 44.19 μm) from group 3 in R1DS1N Negative USAF test target, (10) N-BK7 Ø1" bi-convex lens refractive lens with a focal length ($f = 5$ cm), (11) ID20/M standard iris, Ø20.0 mm max aperture, (12) BS031 50:50 non-polarizing beam splitter cube, (13) Exulus HD2 SLM with 1920 × 1200 pixels and having a pixel size of 8 μm, (14) Ø1" linear polarizer with N-BK7 windows, (15) FLH633-5 band pass filter with a wavelength of λ = 633 nm and a bandwidth of Δλ = 5 nm and (16) Zelux CS165MU/M monochrome image sensor with 1440 × 1080 pixels and having a pixel size of ~3.5 µm.

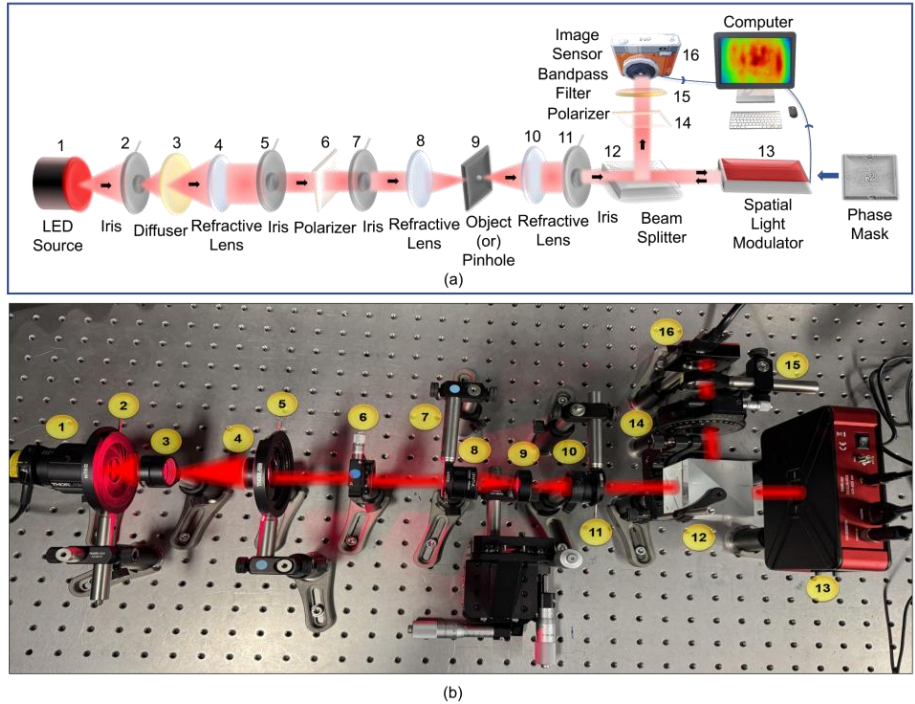

**Figure 2.** (a) Schematic and (b) snapshot of the optical experimental configuration. (1) LED (λ = 660 nm), (2) iris, (3) diffuser, (4) refractive lens ($f$ = 7.5 cm), (5) iris, (6) polarizer, (7) iris, (8) refractive lens ($f$ = 5 cm), (9) object/pinhole, (10) refractive lens ($f$ = 5 cm), (11) iris, (12) beam splitter, (13) SLM, (14) polarizer, (15) bandpass filter, (16) monochrome image sensor.

The diameter of the light beam from the LED is first controlled by an iris and then incident on a diffuser. The diffused light is collected by a refractive lens ($f$ = 7.5 cm) and passed through a linear polarizer, which is oriented along the active axis of the SLM. Irises are placed before and after the polarizer to block out-of-focus stray light from entering the recording unit. The object/pinhole is critically illuminated by a refractive lens ($f$ = 5 cm), and the light beam from the object/pinhole is collected by another refractive lens ($f$ = 5 cm). An iris is used to control the beam size such that the light beam from the lens enters the beam splitter and is incident exactly at the central region of the SLM. The phase masks are displayed on the SLM one after the other, and the holograms are captured by the monochrome image sensor. The second linear polarizer is also oriented along the active axis of the SLM. A bandpass filter is used to enhance the fringe visibility of the recorded holograms. The imaging distance from the SLM to the image sensor is $z_h$ = 20 cm. Four object planes were selected to demonstrate the core concept of PEAR-FINCH, as shown in Figure 3(a). Plane 1 is designated the reference plane. Planes 2, 3, and 4 are located at axial distances of 3 mm, 6 mm, and 12 mm, respectively, from Plane 1. A diffractive lens with a focal length of ($f$ = 20 cm), shown in Figure 3(b), is used to record direct images of the objects. In this study, the digits '2' and '4' are considered Object 1 and Object 2, respectively.

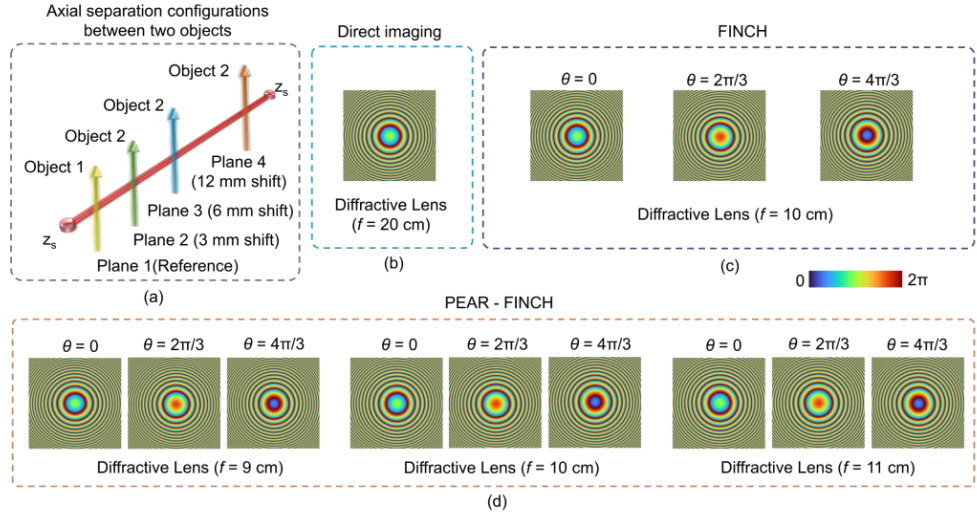

**Figure 3.** (a) Axial separation configurations between two objects. (b) Phase mask for DI – diffractive lens ($f = 20$ cm). (c) Phase masks for FINCH – diffractive lens ($f = 10$ cm) with phase shifts of $\theta = 0$, $2\pi/3$, and $4\pi/3$. (d) Phase masks for the PEAR-FINCH diffractive lens ($f = 9$, 10, and 11 cm) with phase shifts of $\theta = 0$, $2\pi/3$, and $4\pi/3$.

As a preliminary step, the focal depth of the DI method was tested when the object was located at different planes of interest. The DI results are shown in Figure 4. When both objects are placed on plane 1, they appear in perfect focus, as shown in Figure 4(a). When Object 1 remained at plane 1 and Object 2 was shifted to plane 2, Object 2 became blurred, as shown in Figure 4(b). As Object 2 was further moved to planes 3 and 4, the degree of blurring increased, as shown in Figures 4(c) and 4(d), respectively. These observations indicate that, in regular DI, object information is resolved only when the object lies precisely at the focal plane. Even slight axial displacement results in a blur, and the object information is not resolved because of the short DOF.

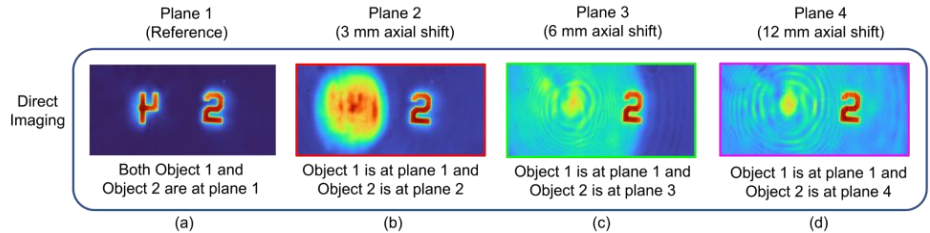

**Figure 4. DI results.** (a) Plane 1 (Reference). (b) Plane 2 (3 mm axial shift). (c) Plane 3 (6 mm axial shift). (d) Plane 4 (12 mm axial shift).

In the next step, the DOF in FINCH was evaluated. To perform this experiment, FINCH was implemented using the polarization multiplexing configuration. Accordingly, two polarizers in the experimental setup before and after the SLM are oriented at 45° with respect to the active axis of the SLM. The first polarizer positioned before the SLM ensures that approximately 50% of the incident light beam is modulated by the diffractive lens displayed on the SLM, whereas the remaining light beam remains unmodulated. As the second polarizer, positioned after the SLM, is also aligned at 45° relative to the SLM's active axis, this configuration enables interference between the modulated and unmodulated light beams, which is essential for FINCH hologram formation. For demonstration, initially, Object 1 was placed at plane 1, whereas Object 2 was placed at plane 2 and then at planes 3 and 4. The phase masks shown in Figure 3(c) were sequentially displayed on the SLM, and the corresponding phase-shifted OHs were recorded.

The 3D imaging results of FINCH are shown in Figure 5. The phase-shifted OHs corresponding to phase shifts of $\theta = 0$, $2\pi/3$, and $4\pi/3$, along with the magnitude and phase of the resulting complex OHs for plane1 + plane 2, plane 1 + plane 3, and plane 1 + plane 4 for a diffractive lens of $f = 10$ cm, are shown in Figure 5(a). The reconstruction results by back propagation to $z_r = 10$ cm, for plane 1 + plane 2 (outlined in red), plane 1 + plane 3 (outlined in green), and plane 1 + plane 4 (outlined in magenta) are shown in Figure 5(b). In all three reconstructions, the reconstruction distance was kept constant, and it corresponds to plane 1 ($z_r = 10$ cm). In the first reconstruction result (outlined in red), both Object 1 and Object 2 appear in focus, even though Object 2 was at plane 2 (3 mm axially located away from plane 1). This demonstrates the high DOF provided by FINCH. When Object 2 was placed at plane 3 (6 mm axially located away from plane 1), it began to appear blurred, as shown in the second reconstruction result (outlined in green). When shifting further to plane 4 (12 mm axially located away from plane 1), Object 2 became significantly blurred, as seen in the third reconstruction result (outlined in magenta). In both of these previous two cases, Object 1 remained in perfect focus, as the reconstruction distance matched its location, where Object 2 was blurred and could not be resolved, as it lies at planes 3 and 4, which are outside the effective DOF region of FINCH.

The goal is to reconstruct the image of Object 2, which lies on plane 3 (6 mm axially away from plane 1), using the proposed PEAR-FINCH method. In PEAR-FINCH, the optical configuration is modified by varying the focal length of the diffractive lenses displayed on the SLM, and corresponding holograms are recorded. In addition to the OH recorded at planes 1 and using three diffractive phase-shifted lenses with a focal length of $f = 10$ cm, six additional phase-shifted OHs were recorded using diffractive lenses with focal lengths of $f = 9$ cm and $f = 11$ cm, as shown in Figure 3(d). In these modified optical configurations, the two interfering beams do not overlap perfectly, and each configuration produces a distinct axial intensity distribution, reconstruction distance, and DOF. The SOH is generated by summing all the complex OHs, exhibiting an extended DOF due to the combination of multiple distinct axial distributions.

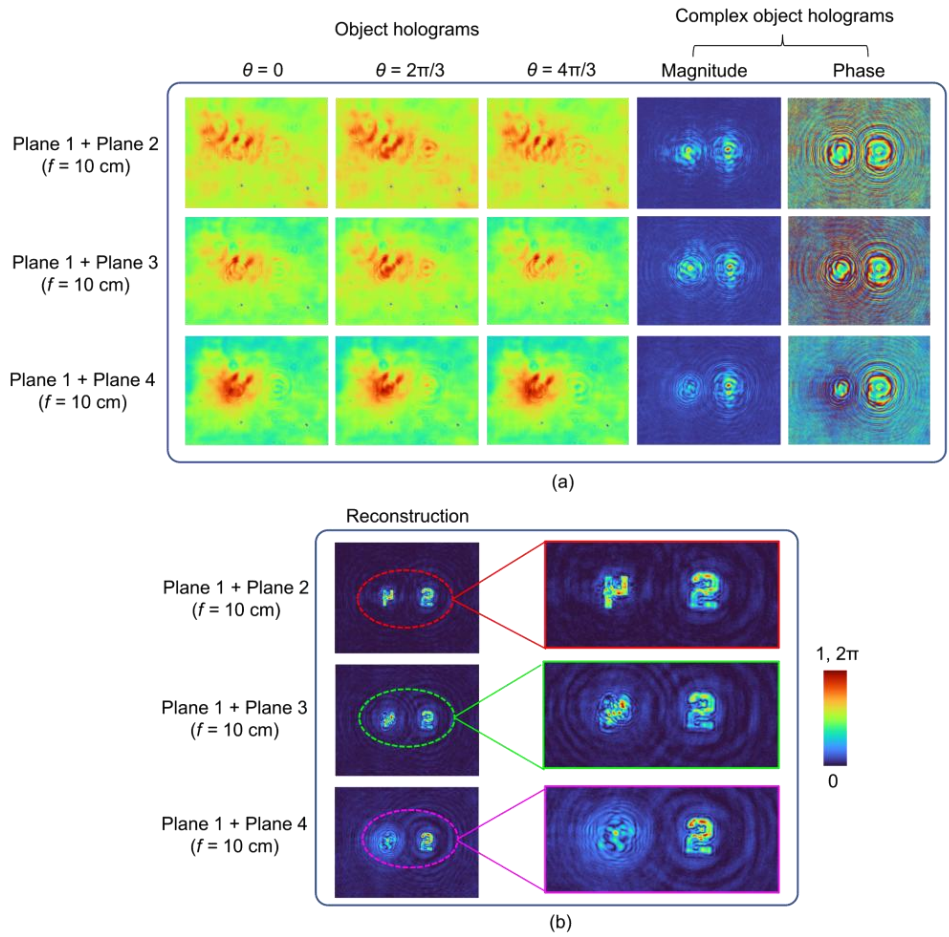

**Figure 5. FINCH 3D imaging results**. (a) Phase-shifted OHs corresponding to phase shifts of $\theta = 0$, $2\pi/3$, and $4\pi/3$, the magnitude and phase of the resulting complex OHs for plane1 + plane 2, plane 1 + plane 3, and plane 1 + plane 4 with diffractive lens of $f = 10$ cm. (b) The reconstruction results by back propagation to $z_r = 10$ cm, for plane 1 + plane 2 (outlined in red), plane 1 + plane 3 (outlined in green), and plane 1 + plane 4 (outlined in magenta).

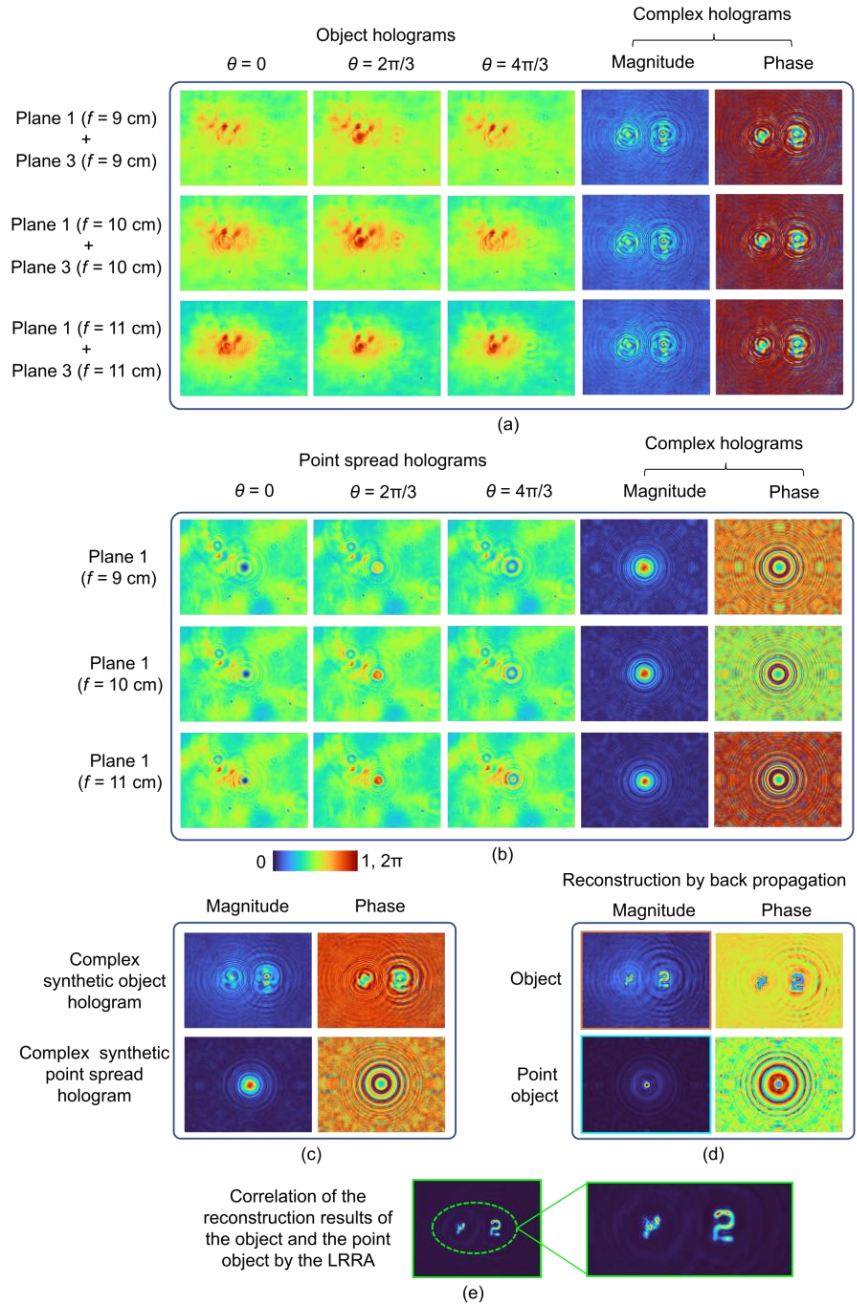

**Figure 6 PEAR-FINCH 3D imaging results**. (a) Phase-shifted OHs corresponding to phase shifts of $\theta = 0$, $2\pi/3$, and $4\pi/3$, the magnitude and phase of the complex OHs for plane 1 ($f = 9$ cm) + plane 3 ($f = 9$ cm), plane 1 ($f = 10$ cm) + plane 3 ($f = 10$ cm), and plane 1 ($f = 11$ cm) + plane 3 ($f = 11$ cm). (b) Phase-shifted PSHs corresponding to phase shifts of $\theta = 0$, $2\pi/3$, and $4\pi/3$, the magnitude and phase of the complex PSHs for plane 1 ($f = 9, 10,$ and $11$ cm). (c) The magnitude and phase of the complex SOH and SPSH. (d) The magnitude (outlined in orange) and phase of the reconstructed object by back-propagation; the magnitude (outlined in cyan) and phase of the reconstructed point object by back-propagation. (e) Correlation of the reconstructed object and the point object shown in (d) via LRRA (outlined in green).

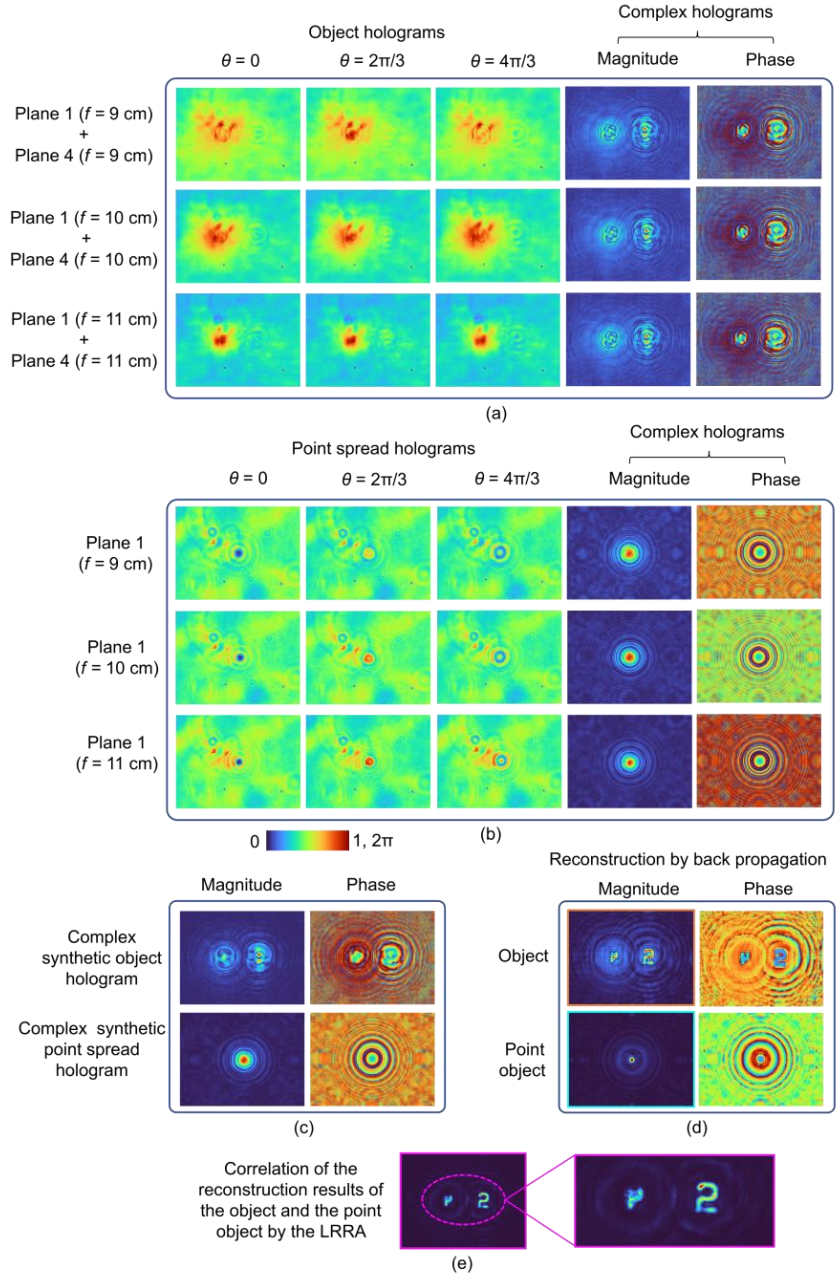

**Figure 7 PEAR-FINCH 3D imaging results**. (a) Phase-shifted OHs corresponding to phase shifts of $\theta = 0$, $2\pi/3$, and $4\pi/3$, the magnitude and phase of the complex OHs for plane 1 ($f = 9$ cm) + plane 4 ($f = 9$ cm), plane 1 ($f = 10$ cm) + plane 4 ($f = 10$ cm), and plane 1 ($f = 11$ cm) + plane 4 ($f = 11$ cm). (b) Phase-shifted PSHs corresponding to phase shifts of $\theta = 0$, $2\pi/3$, and $4\pi/3$, the magnitude and phase of the complex PSHs for plane 1 ($f = 9$, 10, and 11 cm). (c) The magnitude and phase of the complex SOH and SPSH. (d) The magnitude (outlined in orange) and phase of the reconstructed object by back-propagation; the magnitude (outlined in cyan) and phase of the reconstructed point object by back-propagation. (e) Correlation of the reconstructed object and the point object shown in (d) via LRRA (outlined in magenta).

Upon reconstructing the SOH by back-propagation to the reconstruction distance corresponding to plane 1 ($z_r = 10$ cm), both objects were in perfect focus. Notably, the reconstructed image exhibited a low SNR because the SOH contains contributions from multiple OHs. As a result, the reconstruction to a specific distance includes both sharp and blurred reconstructions; when combined, it leads to a low SNR. To improve the SNR, the SPSH is obtained under identical conditions but at only one plane, which is plane 1, with different conditions using diffractive lenses with focal lengths of $f = 9$, 10, and 11 cm. A cross-correlation is then performed between the reconstructed images of the object and the point object using the LRRA. This process enhances the SNR and improves the final reconstruction image quality. The 3D imaging results of PEAR-FINCH are shown in Figure 6. The phase-shifted OHs corresponding to phase shifts of $\theta = 0$, $2\pi/3$, and $4\pi/3$, the magnitude and phase of the complex OHs for plane 1 ($f = 9$ cm) + plane 3 ($f = 9$ cm), plane 1 ($f = 10$ cm) + plane 3 ($f = 10$ cm), and plane 1 ($f = 11$ cm) + plane 3 ($f = 11$ cm) are shown in Figure 6(a). The phase-shifted PSHs corresponding to phase shifts of $\theta = 0$, $2\pi/3$, and $4\pi/3$, the magnitude and phase of the complex PSHs for plane 1 for $f = 9$, 10, and 11 cm are shown in Figure 6(b). The magnitude and phase of the complex SOH and SPSH are shown in Figure 6(c). The magnitude (outlined in orange) and phase of the reconstruction of the object by back-propagation to plane 1 ($z_r = 10$ cm); the magnitude (outlined in cyan) and phase of the reconstruction of the point object by back-propagation to plane 1 ($z_r = 10$ cm) are shown in Figure 6(d). The magnitude of the reconstruction (outlined in orange) indicates that both Object 1 and Object 2 are in focus, which confirms that the focal depth is extended further by the PEAR-FINCH method. The final reconstruction result (outlined in green) with an improved SNR is obtained by processing the reconstructed object and the reconstructed point object shown in Figure 6(d) by LRRA, as is shown in Figure 6(e). The reconstruction parameters, the number of iterations, $\alpha$, and $\beta$, used in LRRA are 3, 0.84, and 1, respectively.

The same procedure is applied to retrieve the information of Object 2, which lies on plane 4 (12 mm axially located away from Plane 1), using the PEAR-FINCH method. In addition to the OH recorded at planes 1 and 4 using three-phase-shifted diffractive lens masks with a focal length of $f = 10$ cm, six additional phase-shifted OHs were recorded using diffractive lens masks with focal lengths of $f = 9$ cm and $f = 11$ cm, as shown in Figure 3(d). The SPSH is obtained at plane 1, with different conditions using diffractive lenses with focal lengths of $f = 9$, 10, and 11 cm are used again to improve the SNR for this case. The 3D imaging results of PEAR-FINCH are shown in Figure 7. The phase-shifted OHs corresponding to phase shifts of $\theta = 0$, $2\pi/3$, and $4\pi/3$, the magnitude and phase of the complex OHs for plane 1 ($f = 9$ cm) + plane 4 ($f = 9$ cm), plane 1 ($f = 10$ cm) + plane 4 ($f = 10$ cm), and plane 1 ($f = 11$ cm) + plane 4 ($f = 11$ cm) are shown in Figure 7(a). The phase-shifted PSHs corresponding to phase shifts of $\theta = 0$, $2\pi/3$, and $4\pi/3$, the magnitude and phase of the complex PSHs for plane 1 with focal lengths $f = 9$, 10, and 11 cm are shown in Figure 7(b). The magnitude and phase of the complex SOH and SPSH are shown in Figure 7(c). The magnitude (outlined in orange) and phase of the reconstructed object by back-propagation to plane 1 ($z_r = 10$ cm); the magnitude (outlined in cyan) and phase of the reconstructed point object by back-propagation to plane 1 ($z_r = 10$ cm) are shown in Figure 7(d). The magnitude of the reconstruction (outlined in orange) indicates that both Object 1 and Object 2 are in focus, which confirms that the focal depth is extended even more as Object 2 is located at the farthest point from plane 1 by the PEAR-FINCH method. The final reconstruction result (outlined in magenta) with an improved SNR is obtained by processing the reconstructed object and the reconstructed point object shown in Figure 7(d) by LRRA, as is shown in Figure 7(e). The reconstruction parameters, the number of iterations, $\alpha$, and $\beta$, used in LRRA are 3, 0.99, and 1, respectively.

The experimental results obtained using DI, FINCH, and the proposed PEAR-FINCH method are compared in Figure 8(a). The PEAR-FINCH method demonstrates superior performance, as both Object 1 and Object 2 remain in perfect focus even when Object 2 is placed at planes 3 and 4, indicating a significantly extended DOF compared with FINCH and DI. To quantitatively validate this performance of PEAR-FINCH, the structural similarity index measure (SSIM) was first calculated for Object 2 when it is placed at planes 3 and 4 for all three methods. The corresponding SSIM bar plot is shown in Figure 8(b). Compared to DI and FINCH, PEAR-FINCH yields a higher SSIM, confirming better preservation of the structural information of the Object 2 at larger axial separations. In the next step, image entropy metrics over the entire image plane were calculated for the cases where Object 1 is placed at plane 1 and Object 2 is placed at plane 3, and Object 1 is placed at plane 1 and Object 2 is placed at plane 4 for all three methods. The corresponding entropy bar plot is shown in Figure 8(c). Compared with DI and FINCH, PEAR-FINCH achieves a lower entropy, confirming improved SNR.

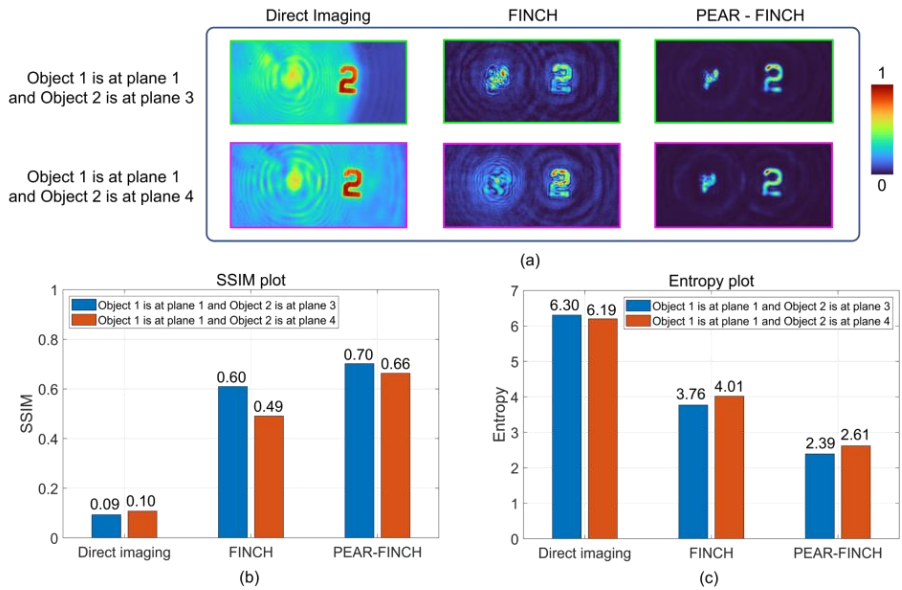

**Figure 8.** (a) Comparison results of DI, FINCH, and PEAR-FINCH for: Object 1 is at plane 1 and Object 2 is at plane 3 (outlined in green); Object 1 is at plane 1 and Object 2 is at plane 4 (outlined in magenta). (b) SSIM bar plots of DI, FINCH, and PEAR-FINCH for Object 2 at plane 3 and Object 2 at plane 4. (c) entropy bar plots of DI, FINCH, and PEAR-FINCH for Object 1 at plane 1 and Object 2 at plane 3, and Object 1 at plane 1 and Object 2 at plane 4.

## 4. Summary and Conclusion

In this study, a new method called PEAR-FINCH is proposed and experimentally demonstrated to extend the DOF of FINCH post-recording on demand for the first time. Among the compared methods, DI has the shortest DOF, FINCH has a longer DOF, and PEAR-FINCH has the longest DOF. In PEAR-FINCH, this is achieved by recording a library of phase-shifted OHs with unique axial configurations generated by varying the focal lengths of diffractive lenses displayed on the SLM and then combining with the FINCH OH to form the SOH, which has a long DOF. The DOF of the SOH depends upon the number of summed holograms. The reconstruction of SOH by backpropagation yields focused images of the objects at different depths, as shown in Figures 6(d) and 7(d), outlined in orange. However, since the reconstruction distance corresponds to a particular distance, there are both sharp and blurred reconstructions from the different holograms in the SOH, which are combined, leading to a lower SNR. To improve the SNR, the PSH is recorded in the same manner but for a single depth, with unique axial configurations. The SPSH is then generated and reconstructed to the corresponding distance. Using the LRRA, the reconstruction of object and point object achieved by backpropagation is cross-correlated, which significantly enhances the SNR, as shown in Figure 6(e), outlined in green, and Figure 7(e), outlined in magenta. The experimental results demonstrate the effectiveness of the PEAR-FINCH method in extending the DOF compared with DI and FINCH. Quantitative analysis using SSIM and entropy confirms that PEAR-FINCH outperforms both DI and FINCH when the object is placed outside the axial distribution range of DI and FINCH. While PEAR-FINCH offers flexible DOF control post-recording, high LR, and improved SNR, it currently requires multiple camera shots, which increases the recording time. Given its improved performance, simplicity, and compatibility compared with existing DI and FINCH systems, the penalty is not very high. However, our future research will focus on reducing the number of required camera shots, potentially enabling single-shot DOF tuning, and extending the PEAR-FINCH method to be utilized in real-world applications with complex natural object scenes. In conclusion, PEAR-FINCH serves as a novel and practical extension of FINCH, offering an extended DOF. Its robustness and flexibility make it suitable for applications in IDH, microscopy, biomedical imaging, and other fields where extended DOF is essential.

## Acknowledgments

The authors thank the funding agencies, the European Union's Horizon 2020 research and innovation programme Grant Agreement No. 857627 (CIPHR), the Australian Research Council, Grant No. DP240103231 and the Israel Science Foundation (ISF) Grant No. 3306/25 for supporting this research project.

## Declaration

The authors declare no competing interests.

## Data availability statement

Data underlying the results presented in this paper can be obtained from the authors upon reasonable request.